\documentclass[runningheads]{llncs}
\usepackage{setup}

\title{Interactions between social norms and incentive mechanisms in organizations}
\titlerunning{Social norms and incentive mechanisms}

\author{Ravshanbek Khodzhimatov\inst{1}\orcidID{0000-0002-2761-2029} \and
Stephan Leitner \inst{2}\orcidID{0000-0001-6790-4651} \and
Friederike Wall\inst{2}\orcidID{0000-0001-8001-8558}}

\authorrunning{R. Khodzhimatov et al.}

\institute{Digital Age Research Center, University of Klagenfurt, 9020 Klagenfurt, Austria
\email{ravshanbek.khodzhimatov@aau.at}\\
\and
Department of Management Control and Strategic Management, University of Klagenfurt, 9020 Klagenfurt, Austria\\
\email{\{stephan.leitner, friederike.wall\}@aau.at}}

\begin{document}
\maketitle 
\begin{abstract}
We focus on how individual behavior that complies with social norms interferes with performance-based incentive mechanisms in organizations with multiple distributed decision-making agents. We model social norms to emerge from interactions between agents: agents observe other the agents' actions and, from these observations, induce what kind of behavior is socially acceptable. By complying with the induced socially accepted behavior, agents experience utility. Also, agents get utility from a pay-for-performance incentive mechanism. Thus, agents pursue two objectives. 
We place the interaction between social norms and performance-based incentive mechanisms in the complex environment of an organization with distributed decision-makers, in which a set of interdependent tasks is allocated to multiple agents. 
The results suggest that, unless the sets of assigned tasks are highly correlated, complying with emergent socially accepted behavior is detrimental to the organization's performance. However, we find that incentive schemes can help offset the performance loss by applying individual-based incentives in environments with lower task-complexity and team-based incentives in environments with higher task-complexity.







\keywords{Agent-based modeling and simulation  \and $NK$-framework \and emergence \and socially accepted behavior}

\end{abstract}
\section{INTRODUCTION}
\label{sec:intro}

Norms are defined as behavior that is common within a society or as rules that are aimed at maintaining specific patterns of behavior which are acceptable to (the majority) of a society \cite{morris19}. In line with this definition, Sen and Airiau \cite{sen07} stress that norms facilitate coordination -- they refer to Lewis \cite{lewis69} who argues: \textit{"Everyone conforms, everyone expects others to conform, and everyone has good reason to conform because conforming is in each person's best interest when everyone else plans to conform"}, and conclude that norms can be interpreted as external correlating signals that promote behavioral coordination.

Despite being a focus of research in many scientific disciplines, a consensus about the ontology of norms has not yet been reached \cite{mellema20}. In this paper, we follow 
the classification introduced by Tuomela et al. \cite{tuomela95}, who distinguish between four types of norms, namely  (i) rule norms, (ii) proper social norms, (iii) moral norms, and (iv) prudential norms (see also \cite{morris19}). 
They argue that the (i) \textit{rule norms} can be either formal or informal. The former are articulated and written with formal sanctions and brought into existence by an authority, the latter are articulated but usually not written down, associated with informal sanctions and brought into existence by group members' mutual implicit or explicit agreement. Morris-Martin et al. \cite{morris19} add that rule norms are often also referred to as laws. 
 With respect to (ii) \textit{social norms}, Tuomela et al. \cite{tuomela95} distinguish between \textit{conventions}, which apply to an entire community, society or social class, and \textit{group-specific norms}, which are specific to one or more groups, but not to the entire society. This understanding of social norms is in line with the definition introduced in Cialdini et al. \cite{cialdini98}, who add that social norms are usually not accompanied by enforcing laws. Mahmoud et al. \cite{mahmoud19} stress that social norms can be interpreted as informal rules and standards which entail what others expect, e.g., in terms of behavior, and have a non-obligatory character. The latter implies that social norms are self-enforcing and that there are often social processes underlying norms that ensure that non-conforming results in a social punishment \cite{axelrod97,sen07}. Thus, obeying social norms is often regarded to be rational due to the threat of social sanctions \cite{elster89}.
Finally, Morris-Martin et al. \cite{morris19} and Tuomela et al. \cite{tuomela95} line out that (iii) \textit{moral norms} are intended to appeal an individual's conscience and (iv) \textit{prudential norms} usually follow the principles of rationality. 

In this paper, we adopt the notion of (ii) \textit{social norms} introduced above. Their presence has been widely recognized in the academic literature. The field of multi-agent systems is, for example, concerned with the emergence of social norms and their enforcement in agent societies. Recent reviews of research on norms in multi-agent systems are provided by Morris-Martin et al. \cite{morris19} and Alechina et al. \cite{alechina18}. Cranefield et al. \cite{cranefield17} line out that the way norms are included in decision making algorithms needs to be explicitly formulated by the designer, while for human agents, norms (and values) are highly entrenched in the decision making process. This is in line with Kohlberg \cite{kohlberg84}, who argues that individuals have an endogenous preference to conform to the behavior of their peers, which is why social norms play a central role in a multiplicity of contexts in which humans interact and make decisions, such as decisions between different courses of action in organizations or in politics \cite{sen07}.

We apply social norms to the context of organizations which consist of collaborative and distributed decision makers and focus on the interaction between emergent social norms (at the level of individuals) and performance-based incentives, the behavioral implications of this interaction, and its consequences for the performance of the overall system.\footnote{Note that along with social norms at the individual level, previous research also addresses social norms at the level of organizations: Dowling et al. \cite{dowling75}, for example, conceptualize organizational legitimacy as congruence between the social values associated with an organization's action and the norms of acceptable behavior in the social system of the organization. This paper, however, focuses on social norms \textit{within} an organization.} By doing so, we focus on social norms which emerge from past decisions of fellow agents within an organization \cite{cialdini90}.

The remainder of this paper is organized as follows: Sec. \ref{sec:litreview} reviews the research on social norms, Sec. \ref{sec:model} describes the structure and methodology we use to model the simulation of organizational environment with emergent social norms and varying performance-based incentives, Sec. \ref{sec:results} elaborates on results and findings, and Sec. \ref{sec:conclusion} concludes this paper.

\section{RELATED WORK}
\label{sec:litreview}

Considering social norms as key-factors which drive individual behavior is not a new issue in research. Early work on this topic goes back to social approval of individual behavior \cite{hayek67,smith1822}. A recent survey of interactions between social norms and incentives is provided by Festre \cite{festre10}, who argues that social norms are supplementary motives to self-interest, which is commonly assumed for economic agents. He presents the following two empirical findings to support this assertion: (i) a large proportion of Americans do not apply for welfare programs, even when they are eligible \cite{moffitt83}, and (ii) in donations to charity, whether the list of contributors is published or not, has an effect on the total amount donated \cite{andreoni00,glazer96}. Festre \cite{festre10} claims that the reason for this behaviour can be traced back to social norms.\footnote{For extensive discussions on the role of social norms in behavioral control, the reader is also referred to \cite{kubler01}, \cite{vendrik03}, and most recently \cite{lieberman19} and the literature cited in these studies.} 

Festre \cite{festre10} reviews current studies on social norms in economic theory and concludes that there might be two explanations for social norms as a driver of individual behavior: (i) the individual desire for conformity and (ii) positive externalities. For (i) the individual desire for conformity, she argues that individuals care about their social status (e.g., in terms of popularity or respect) and therefore want to conform to social norms. This explanation is in line with previous studies \cite{bernheim94,veblen73}. With respect to (ii) positive externalities, Festre \cite{festre10} refers to Coleman \cite{coleman88}, who lines out that situations, in which the same outcome satisfies the interests of others, enforce social norms. Consequently, since everyone has incentives to reward others for working towards this outcome, all individuals have two sources of utility: the reward for the effort one made towards the outcome (i.e., the incentives), and the rewards provided by others for helping to achieve that outcome (in terms of social approval). This argumentation is also in line with Janssen et al. \cite{janssen04} and Huck et al. \cite{huck12}, who argue that individual behavior is driven by multiple forces and that interactions may exist among these forces (e.g., in terms of reinforcement or weakening).  
Festre \cite{festre10} adds that previous research has shortcomings in the way it deals with behavioral responses to norms or changes in norms, and that the interaction between \textit{endogenous} social norms and incentives should be further addressed. The latter is also in line with K{\" u}bler \cite{kubler01}, who argues that social norms have been considered as being exogenous (and, thus, not emergent) for (too) long. Some work in the field of psychology addresses social norms, but puts the focus on the emergence in the sense of learning which type of behavior is socially approved: Paluck et al. \cite{paluck12}, for example, are concerned with evolving norms in the context of social networks in schools and Ehrhart et al. \cite{ehrhart04} address citizenship behavior in organizational units. Previous research in the field of economics has, amongst others, addressed how performance-based incentives can change the meaning of following a social norm \cite{huck12,kubler01}, and how incentive framing and social norms interact with respect to behavioral implications \cite{lieberman19}. Moreover, previous research has found that, under specific circumstances, monetary incentives can crowd out incentives provided by intrinsic factors, such as social norms \cite{benabou03,janssen04}. 

We implement emergent social norms in the sense of Cialdini et al. \cite{cialdini90}, who state that social norms emerge from the shared knowledge about the past behavior of peers and are determined by the strength of social interactions and the similarity of decisions, regardless of the impact of the norms on the outcomes. Thus, we focus on the behavior that is \textit{normal} in the population, rather than what is declared (morally or otherwise) to be a desired behavior \cite{ajzen91,cialdini90}. This allows us to model the social norms solely as an emergent phenomenon without imposing the desirability qualities to particular actions.

Fischer and Huddart \cite{fischer08} similarly acknowledge that social norms can emerge endogenously: they argue that there is a complex relationship between individual behavior and social norms within an organization, as they are mutually dependent. If an agent's peers are members of the same organization, individual behavior determines the organization's social norms, which, in turn, influence individual behavior. They acknowledge, however, that individual behavior of the members of an organization is affected not only by social norms but also by other means of behavioral control, such as incentive systems. They conclude that social norms (i) emerge endogenously within organizations from the individual behavior of the organization's members and their interaction, and (ii) might be endogenously affected by choices related to organizational design elements. They explicitly point out that further investigation of the interaction between social norms and incentives is required. This is where we place our research: we study how social norms affect the performance in organizations with collaborative and distributed decision makers and how they interact with performance-based incentive mechanisms.
\section{MODEL}
\label{sec:model}

This section introduces the model of a stylized organization which is implemented as a collective of $P$ agents facing a complex task. The task environment is based on the $NK$-framework \cite{kauffman89,wall21,leitner14}. Agents face the dilemma of pursuing two objectives simultaneously, namely, to conform to emergent social norms and to maximize their individual (performance-based) utility. We model agents to employ the approach of goal programming to dissolve this dilemma and observe how interactions between social norms and (performance-based) incentives affect the organization's performance for $t=\{1,2,\dots,T\}$ periods. The task environment, in which the organization operates, is introduced in Sec. \ref{sec:design}, while Secs. \ref{sec:agents} and \ref{sec:social} characterize the agents and describe how social norms emerge, respectively. Section \ref{sec:discovering} describes the agents' search for better performing solutions to the decision problem and the approach of goal programming is introduced in Sec. \ref{sec:multiobj}. Finally, Sec. \ref{sec:process} provides an overview of the sequence of events during simulation runs.

\subsection{Task environment}
\label{sec:design}

\begin{figure}[!tb]
    \centering
    \begin{minipage}{0.45\linewidth}
        \centering
        \includegraphics[width=0.95\linewidth]{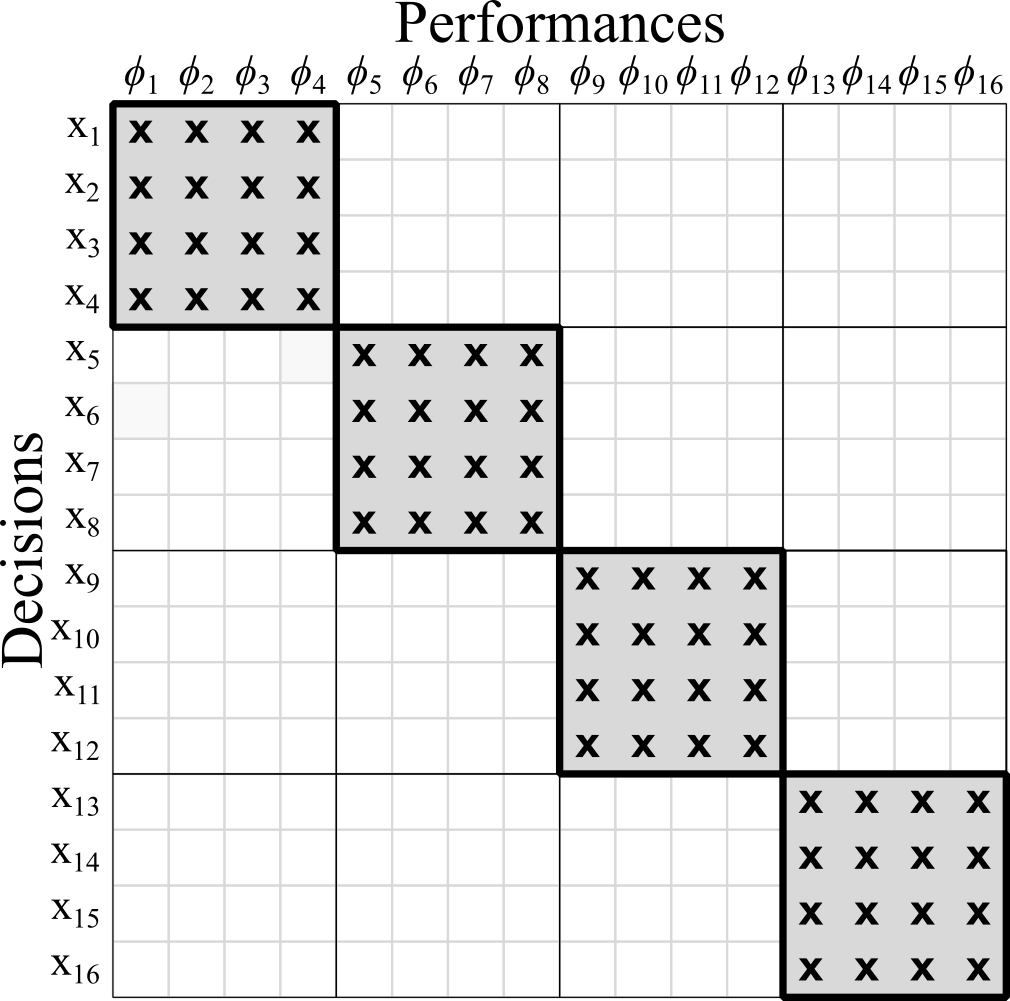}
        \captionsetup{justification=centering}
        \caption*{Internal\\{\small ($K=3,C=S=0$)}}
    \end{minipage}
    \begin{minipage}{0.45\linewidth}
        \centering
        \includegraphics[width=0.95\linewidth]{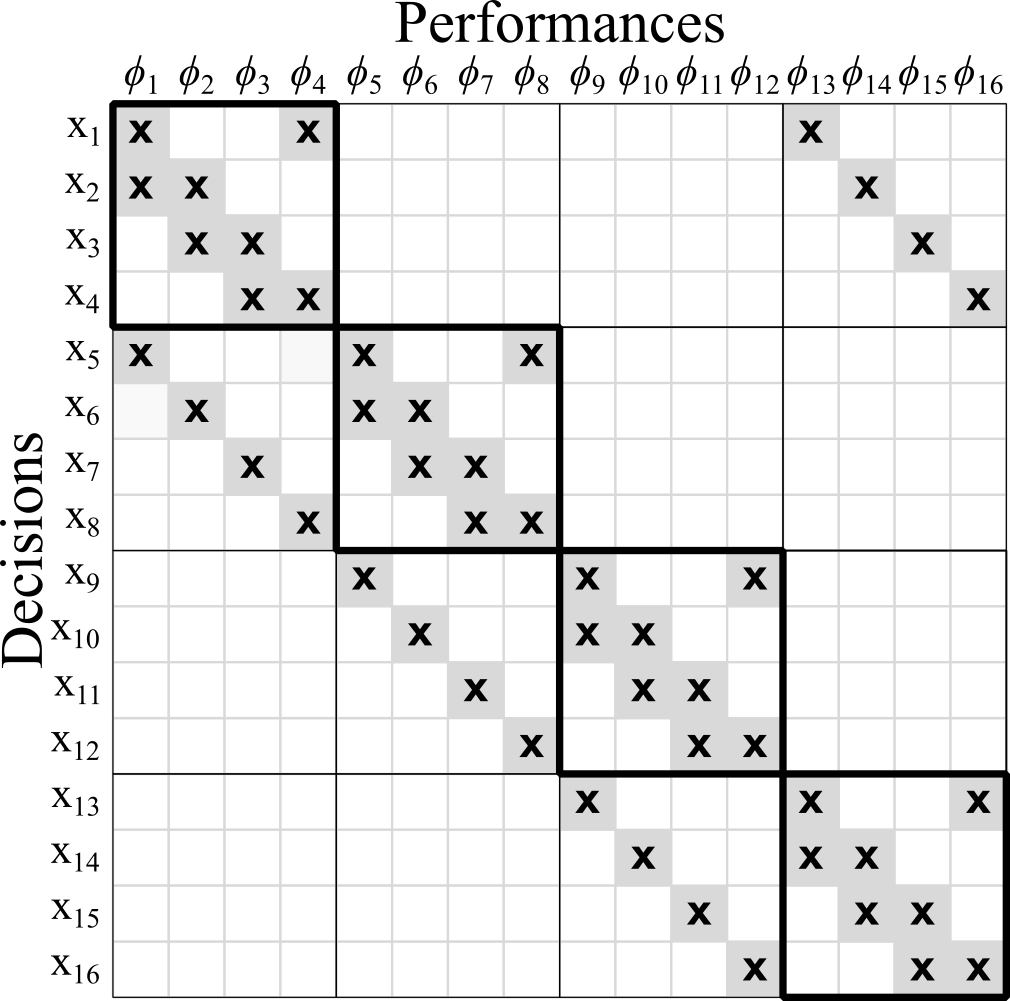}
        \captionsetup{justification=centering}
        \caption*{Low\\{\small ($K=C=S=1$)}}
    \end{minipage}
    
    \vspace{0.5em}
    
    \begin{minipage}{0.45\linewidth}
        \centering
        \includegraphics[width=0.95\linewidth]{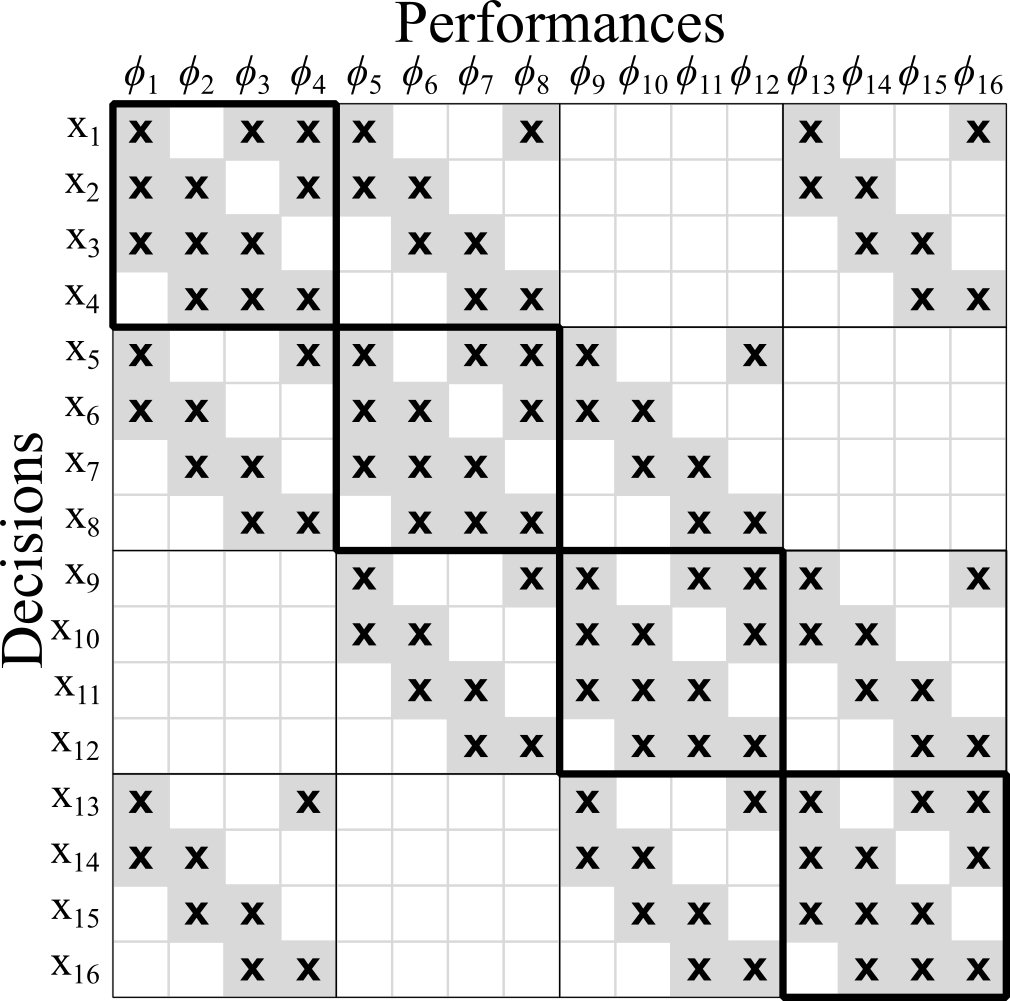}
        \captionsetup{justification=centering}
        \caption*{Moderate\\{\small ($K=C=S=2$)}}
    \end{minipage}
    \begin{minipage}{0.45\linewidth}
        \centering
        \includegraphics[width=0.95\linewidth]{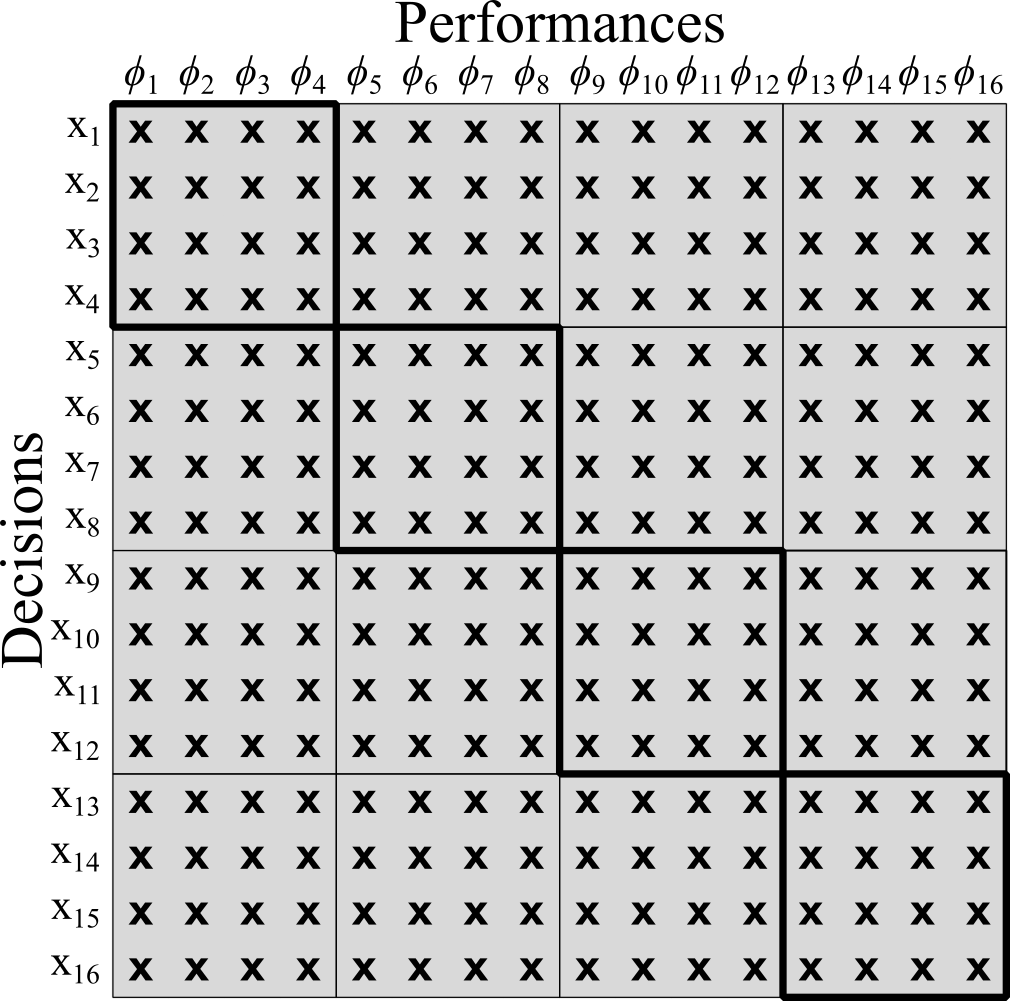}
        \captionsetup{justification=centering}
        \caption*{High\\{\small ($K=S=3,C=4$)}}
    \end{minipage}
    \caption{Stylized interdependence structures with $M = 16$ tasks equally assigned to $P=4$ agents for different levels of complexity. The crossed cells indicate inter-dependencies as follows: let $(i,j)$ be coordinates of a crossed cell in row-column order, then performance contribution $\phi(x_{i})$ depends on decision $x_j$.}
    \label{fig:interactions}
\end{figure}

We model an organization that faces a complex decision problem that is expressed as the set of $M$ binary choices. The decision problem is segmented into sub-problems which are allocated to $P$ agents, so that each agent faces an $N$-dimensional sub-problem. We denote the organization's decision problem by an $M$-dimensional bitstring $\mathbf{x} = (x_1, x_2, \dots, x_{M})$, where $M=N\cdot P$, bits $x_i$ represent single tasks, and $x_i \in \{ 0, 1 \}$ for $i \in \{1,2,...,M\}$. Without loss of generality, we model tasks to be assigned to agents sequentially, such that agent $1$ is responsible for tasks 1-4, agent $2$ -- for tasks 5-8, and so forth. Formally, agent $p \in \{1,2, \dots, P$\} is responsible for the following vector of tasks: 
\begin{equation}
    \mathbf{x}^p = (x^p_1,\dots,x^p_N) = \left( x_{N\cdot(p - 1)+1}, \dots, x_{N\cdot p} \right)
    \label{eq:tasks}
\end{equation}
Every task $x_i$ for $i \in \{1,2,\dots,M\}$ is associated with a uniformly distributed performance contribution $\phi(x_i) \sim U(0,1)$. The decision problem is \textit{complex} in that the performance contribution $\phi(x_i)$, might be affected not only by the decision $x_i$, but also by decisions $x_j$, where $j \neq i $. We differentiate between two types of such inter-dependencies: (a) \textit{internal} inter-dependencies within $\mathbf{x}^p$, in which interdependence exists between the tasks assigned to agent \(p\), and (b) \textit{external} inter-dependencies between $\mathbf{x}^p$ and $\mathbf{x}^q$, in which interdependence exists between the tasks assigned to agents $p$ and $q$, where $p \neq q$. We control inter-dependencies by parameters $K,C,S$, so that every task interacts with exactly $K$ other tasks internally and $C$ tasks assigned to $S$ other agents externally \cite{kauffman91}. Figure \ref{fig:interactions} illustrates four stylized interaction structures considered in this paper. The figure features $M = 16$ tasks equally assigned to $P=4$ employees for different levels of complexity.

Based on the structure outlined above, we can formally describe the performance contribution of decision $x_i$ as follows:
\begin{equation}
    \phi (x_i) = \phi (x_i ,
        \underbrace{x_{i_1},...,x_{i_K}}_{\substack{K\textbf{ internal}\\\text{interdependencies}}}, \underbrace{x_{i_{K+1}} ,...,x_{i_{K + C \cdot S}}}_{\substack{C \cdot S \textbf{ external}\\ \text{interdependencies}}}),
    \label{eq:payoff}
\end{equation}
where $\{i_1, \dots , i_{K+C\cdot S} \} \subset \{1,\dots,M\} \backslash i$, and the parameters satisfy $0 \leq K < N$, $0 \leq C \leq N$, and $0 \leq S < P$. Using Eq. \ref{eq:payoff}, we compute performance landscapes for all agents. We indicate time steps by $t \in \{ 1,2,\ldots,T\}$. Let $\mathbf{x}^p_t$ and $\mathbf{x}_t$ be a vector of decisions of agent $p$ and a vector of decisions of all agents at time $t$, respectively. Then the performance achieved by agent $p$ at time step $t$ is:
\begin{equation}
    \phi_{own} (\mathbf{x}^p_t) = \frac{1}{N} \sum_{x_i \in \mathbf{x}^p_t} \phi(x_i),
    \label{eq:performance-agent}
\end{equation}
and the organization's performance at time step $t$ is:
\begin{equation}
    \phi_{org} (\mathbf{x}_{t}) = \frac{1}{P} \sum_{p=1}^{P} \phi(\mathbf{x}^p_t)~.
    \label{eq:performance-org}
\end{equation}

In order to capture diversity (or similarity) in the sub-problems allocated to agents, we consider the correlations between the performance landscapes using the methodology described in Verel et al. \cite{verel13}. The performance contributions of every set of $N$ tasks assigned to agent $p$ are correlated to the performance contributions of the sets of $N$ tasks assigned to $P-1$ other agents with a constant correlation coefficient $\rho \in [0,1]$. When $\rho=0$ and $\rho=1$, agents operate on perfectly distinct and perfectly identical performance landscapes, respectively.

\subsection{Agents' performance-based incentives}
\label{sec:agents}

The agents' compensation is composed of a fixed and a variable component: without loss of generality, we normalize the former to $0$.  The latter is based on agent $p$'s own performance $\phi_{own}$ (see Eq. \ref{eq:performance-agent}), and the residual performance $\phi_{res}$ resulting from decisions of all other agents. Let $\mathbf{x}^{-p}_t$ be a vector of decisions of all agents other than $p$:
\begin{equation}
    \mathbf{x}^{-p}_t = \{ \mathbf{x}_t^q : q \in \{1,\dots,P\} \backslash p \}
    \label{eq:action-residual}
\end{equation}
Then, the residual performance is defined as the mean of own performances of every agent other than $p$:
\begin{equation}
    \phi_{res} (\mathbf{x}^{-p}_t) = \frac{1}{P-1} \cdot \sum_{\mathbf{x} \in \mathbf{x}^{-p}_t} \phi_{own}(\mathbf{x}),
    \label{eq:performance-residual}
\end{equation}
and agent $p$'s variable compensation component follows the linear incentive scheme\footnote{In our context linear incentives are as efficient as other contracts inducing non-boundary actions. See \cite[p. 1461]{fischer08}.}: 
\begin{equation}
    \phi_{inc}(\mathbf{x}^p_t,\mathbf{x}^{-p}_t) = \alpha \cdot \phi_{own} (\mathbf{x}^p_{t}) + \beta \cdot \phi_{res} (\mathbf{x}^{-p}_t), 
    \label{eq:utility0}
\end{equation}
where $\alpha + \beta = 1$.

\subsection{Social norms}
\label{sec:social}
We implement the emergent social norms using a version of the Social Cognitive Optimization algorithm \cite{xie02}. The algorithm features \textit{social sharing libraries}, where agents share and observe the information (i.e., the previous decisions) which they consider in their decision-making later. In our implementation, every agent has an individual sharing library (as a memory), and the sharing of information happens unidirectionally in directed social networks. Below we explain this algorithm in detail. 

First of all, we differentiate between two types of tasks, namely \textit{private} and \textit{social} tasks. Private tasks are unique to agents, i.e., these tasks cover activities which are in the area of expertise of a specific agent; within the stylized organization captured by our model, only one agent will carry out such a task. In an organization, for example, only the accounting department will be responsible for the accounts payable and the monthly payroll. Social tasks, on the contrary, are types of tasks which (in a similar way) concern all agents. In an organization, every department head will have to make decisions related to their management style, irrespective of the department. In our formulation, private tasks are not relevant to social norms, while social tasks are.

Without loss of generality we use the following convention: let $N_s$ indicate the number of social tasks allocated to each agent. Then the last $N_s$ tasks assigned to agent $p$ are social:

\begin{equation}
    \mathbf{x}^p = 
    ( \underbrace{x^p_1,\dots,x^p_{N-N_s}}_{\text{private tasks}},
    \underbrace{x^p_{N-N_s+1},\dots,x^p_N}_{\text{social tasks}}
    )
\end{equation}

At every time step $t$, agents share the decisions on $N_{s}$ social tasks with $D$ fellow agents in the same organization according to the network structure predefined by the modeler.\footnote{We use the bidirectional \textit{ring network} topology, in which each node is connected to exactly two other nodes with reciprocal unidirectional links, where nodes represent agents and the links represent sharing of information.} Every agent is endowed with a memory $L^p$ in which the decisions on social tasks, made and shared by other agents, are stored. Due to cognitive limitations, the agent's memory is considered to be limited to $T_L$ periods. Once the agents' cognitive capacity is reached, they forget (remove from their memory $L^p$) the oldest information on their fellow agents' decisions on social tasks, i.e., they just remember what was shared in the last $T_L$ periods and forget everything that was shared before. 
Thus, at every time step $t$, agent $p$ gets information about the decisions made on social tasks $\mathbf{x}^q_{soc}$ from $D$ fellow agents $q \in \{p_1,\dots, p_D \} \subseteq \{1,\dots,P\} \backslash p$, and stores it for $T_L$ time steps in memory $L^p$. Social norms do not form in the organization until time period $T_L$.

The extent to which agent $p$'s decision at time $t$, $\mathbf{x}^{p}_{t}$, complies with the emergent social norm is computed as a match rate of the social bits in the memory:
\begin{equation}
    \phi^p_{soc} (\mathbf{x}^p_{t}) = \begin{cases}
        \displaystyle\frac{1}{N_s \cdot |L^p_t|} \sum_{\mathbf{x} \in L^p_t} h(\mathbf{x}^p_{soc}, \mathbf{x}), & t > T_L \\
         0, & t \leq T_L
    \end{cases}
    \label{eq:norms}
\end{equation}
where $|L^p_t|$ is the number of entries in agent $p$'s memory at time $t$ and $h(\mathbf{x},\mathbf{y})$ for two equal-length bitstrings $\mathbf{x}$ and $\mathbf{y}$ of size $J$ is the number of positions at which the corresponding bits are equal:
\begin{equation}
    h \left( \mathbf{x}, \mathbf{y} \right) = \sum^J_{i=1} [ x_i == y_i ]~.
\end{equation}
If the statement inside the bracket is true, it equals $1$, and $0$ otherwise \cite{iversion62}.


\subsection{Discovering new solutions to the decision problem}
\label{sec:discovering}

At time $t$, agent $p$ can observe its own performance in the last period, $\phi_{own} (\mathbf{x}^p_{t-1})$, and the decisions of all agents in the organization in the last period \textit{after} they are implemented, $\mathbf{x}_{t-1}$.

In order to come up with new solutions to their decision problems, agents perform a search in the neighbourhood of $\mathbf{x}_{t-1}$ as follows: agent $p$ randomly switches one decision $x_i \in \mathbf{x}^p$ (from $0$ to $1$, or vice versa), and assumes that other agents will not switch their decisions\footnote{Levinthal \cite{levinthal97} describes situations in which agents switch more than one decision at a time as \textit{long jumps} and states that such scenarios are less likely to occur, as it is hard or risky to change multiple processes simultaneously.}. We denote this vector with one switched element by $\hat{\mathbf{x}}^p_t$. 

Next, the agent has to make a decision whether to stick with the status quo,  $\mathbf{x}^p_t$, or to switch to the newly discovered $\hat{\mathbf{x}}^p_t$. The rule for this decision is described in the next subsection.  

\subsection{Balancing performance-based incentives and social norms and making a decision}
\label{sec:multiobj}

Agents pursue two objectives simultaneously: they aim at maximizing their performance-based incentives formalized in Eq. \ref{eq:utility0} and, at the same time, want to comply with the social norms as formalized in Eq. \ref{eq:norms}. In order to balance these two objectives, agents follow the approach of \textit{goal programming} \cite{charnes57} as described below.

Let $g_{soc}$ and $g_{inc}$ be the goals that agents have for $\phi_{soc}(\mathbf{x}^p_t)$ and $\phi_{inc}(\mathbf{x}^p_t,\mathbf{x}^{-p}_t)$, respectively\footnote{Note that agents are homogeneous with respect to goals and that goals are constant over time.}. Agent $p$ wants to achieve both goals, so that:
\begin{subequations}
\begin{eqnarray}
    \phi_{soc}(\mathbf{x}^p_t) &\geq & g_{soc} \text{, and} \\
    \phi_{inc}(\mathbf{x}^p_t,\mathbf{x}^{-p}_t) & \geq & g_{inc}
\end{eqnarray}
\label{eq:objectivefunction0}
\end{subequations}
Let $d_{soc} (\mathbf{x}^p_t)$ and $d_{inc} (\mathbf{x}^p_t, \mathbf{x}^{-p}_t)$ be the under-achievements of the set of decisions $(\mathbf{x}^p_t, \mathbf{x}^{-p}_t)$ on the goals regarding social norms and performance-based incentives respectively (see Eqs. \ref{eq:utility0} and \ref{eq:norms}):
\begin{align}
    d_{soc} (\mathbf{x}^p_t)                    &= \max \{ g_{soc} - \phi_{soc} (\mathbf{x}^p_t) , 0 \}, \\
    d_{inc} (\mathbf{x}^p_t, \mathbf{x}^{-p}_t) &= \max \{ g_{inc} - \phi_{inc} (\mathbf{x}^p_t, \mathbf{x}^{-p}_t) , 0 \}
\end{align}

As mentioned before, agent $p$ discovers $\hat{\mathbf{x}}^p_t$ -- an alternative configuration to the decision at $t$,  but can only observe what other agents implemented at the previous time period, $\mathbf{x}^{-p}_{t-1}$.
Given $p$'s information, this agent makes the decision to either implement $\hat{\mathbf{x}}^p_t$ or to stick with $\mathbf{x}^p_{t-1}$ at $t$ and chooses $\mathbf{x}^p_{t}$ according to the following rule: 
\begin{equation}
    \mathbf{x}^p_t = \underset{\mathbf{x} \in \{\mathbf{x}^p_{t-1}, \hat{\mathbf{x}}^p_{t}\}}{\arg\min}  w_{soc} \cdot d_{soc} (\mathbf{x}) + w_{inc} \cdot d_{inc} (\mathbf{x},\mathbf{x}^{-p}_{t-1}),
\end{equation}
where $w_{soc}$ and $w_{inc}$ represent the weights for the goal for compliance with the social norms ($g_{soc}$) and goal for performance-based incentives ($g_{inc}$) respectively.

\subsection{Process overview, scheduling and main parameters}
\label{sec:process}
\begin{figure}
	\centering
	\includegraphics[width=\linewidth]{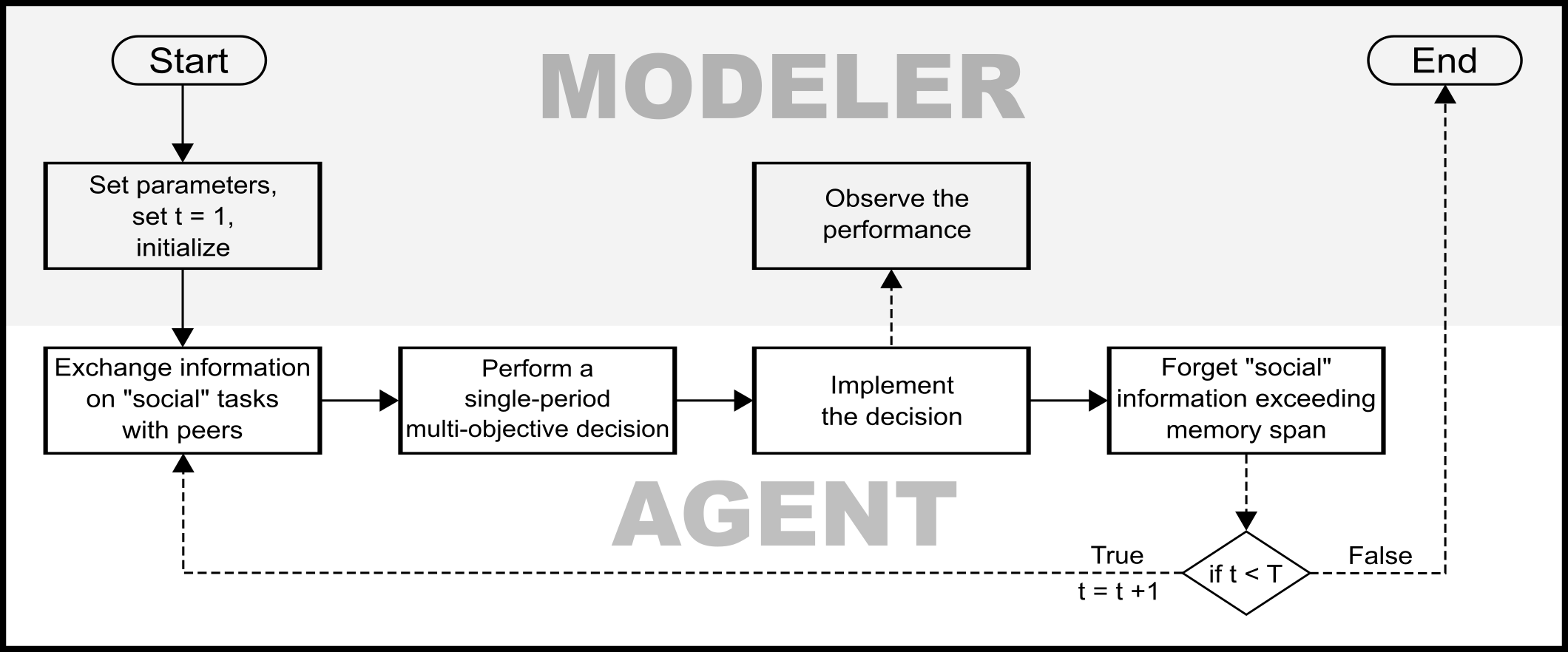}
	\caption{Process overview. Upper actions are performed by the modeler and lower actions are performed by agents.}
	\label{fig:funcflowchart}
\end{figure}

The simulation model has been implemented in Python 3.7.4. Every simulation round starts with the initialization of the agents' performance landscapes, the allocation of tasks to $P=4$ agents\footnote{For reliable results, we generate the entire landscapes before the simulation, which is computationally feasible for $P=4$ given modern RAM sizes. Our sensitivity analyses with simpler models without entire landscapes, suggest that the results also hold for larger population sizes.}, and the generation of an $N$-dimensional bitstring as a starting point of the simulation run (see Sec. \ref{sec:design}). After initialization, agents perform the \textit{hill climbing} search procedure outlined above (see Secs. \ref{sec:discovering} and \ref{sec:multiobj}) and share information regarding their social decisions in their social networks (see Sec. \ref{sec:social}). The observation period $T$, the memory span of the employees $T_L$, and the number of repetitions in a simulation, $R$, are exogenous parameters, whereby the latter is fixed on the basis of the coefficient of variation. Figure \ref{fig:funcflowchart} provides an overview of this process and Tab. \ref{tab:params} summarizes the main parameters used in this paper.

\begin{table}[tb]
    \centering
    \caption{Main parameters}
    \label{tab:params}
    \begin{tabular}{c p{18em} >{\raggedright\arraybackslash}p{8em}}
        Parameter & Description                                 & Value         \\
        \hline
        $M$       & Total number of tasks                       &   16 \\
        $P$       & Number of agents                            & 4             \\
        $N$       & Number of tasks assigned to a single agent  & 4             \\
        $[K,C,S]$ & Internal and external couplings             & $[3,0,0]$, $[1,1,1]$, $[2,2,2]$, $[3,4,3]$    \\                  
        $\rho$    & Pairwise correlation coefficient between tasks assigned to different agents
                                                                & 0.3      \\
        $T_L$     & Memory span of agents                            & 20            \\
        $N_{S}$ & Number of social tasks                        & 2             \\
        $D$       & Level of social connection (network degree) & 2             \\
        $T$       & Observation period                          & 500           \\
        $R$       & Number of simulation runs per scenario      & 300           \\
        $[ g_{inc}, g_{soc}]$ & Goals for performance-based incentives ($\phi_{inc}(\mathbf{x}^p_t,\mathbf{x}^{-p}_t)$) and compliance with the social norms ($\phi_{soc}(\mathbf{x}^p_{t})$)                          & $[1.0,1.0]$   \\
        $[ w_{inc}, w_{soc}]$ & Weights for performance-based incentives $\phi_{inc}(\mathbf{x}^p_t,\mathbf{x}^{-p}_t)$ and compliance with the social norms $\phi_{soc}(\mathbf{x}^p_{t})$                                                           & $[1,0]$, $[0.7,0.3]$, $[0.5,0.5]$ \\
        $[\alpha,\beta]$ & Shares of own and residual performances included in the performance-based incentive scheme                                                                 & $[1,0]$, $[0.75,0.25]$, $[0.5,0.5]$, $[0.25,0.75]$ \\
        \hline
    \end{tabular}
\end{table}


\section{RESULTS}
\label{sec:results}


\subsection{Performance measure}

We indicate the solution (at the system's level) implemented at time step $t$ and simulation run $r\in \{1,\dots,R\}$ by $\mathbf{x}^r_{t}$, and the associated performance by $\phi^r_{org} (\mathbf{x}^r_{t})$ (see Eq. \ref{eq:performance-org}). As the performance landscapes on which agents operate are randomly generated, for every simulation run, we normalize the performances by the maximum performances per landscape to ensure comparability. We indicate then normalized performance achieved by the organization at time step $t$ in simulation run $r$ by 

\begin{equation}
    {\Phi}[r,t] = \frac{\phi^r_{org} (\mathbf{x}^r_t)}{\displaystyle\max_{\mathbf{x} \in [0,1]^M} \{ \phi^r_{org} (\mathbf{x}) \}} 
\end{equation}
We denote the average performance at $t$ by:
\begin{equation}
    \overline{\Phi}[t] = \frac{1}{R} \sum_{r=1}^R \Phi[r,t],
\end{equation}

In Sec. \ref{sec:findings}, we report the \textit{distance to maximum performance} as a performance measure. Note that this measure captures the cumulative distance between the average performances achieved throughout the observation period and the maximum performance attainable (which equals $1$ by construction), and lower (higher) values of the distance indicate higher (lower) performance:

\begin{equation}
    d(\overline{\Phi}) = \sum_{t=1}^T \left( 1 - \overline{\Phi}[t] \right)
\end{equation}

\subsection{Results of the simulation study}
\label{sec:findings}

\begin{figure}[!tb]
	\centering
	\begin{minipage}{0.49\linewidth}
		\includegraphics[width=\linewidth]{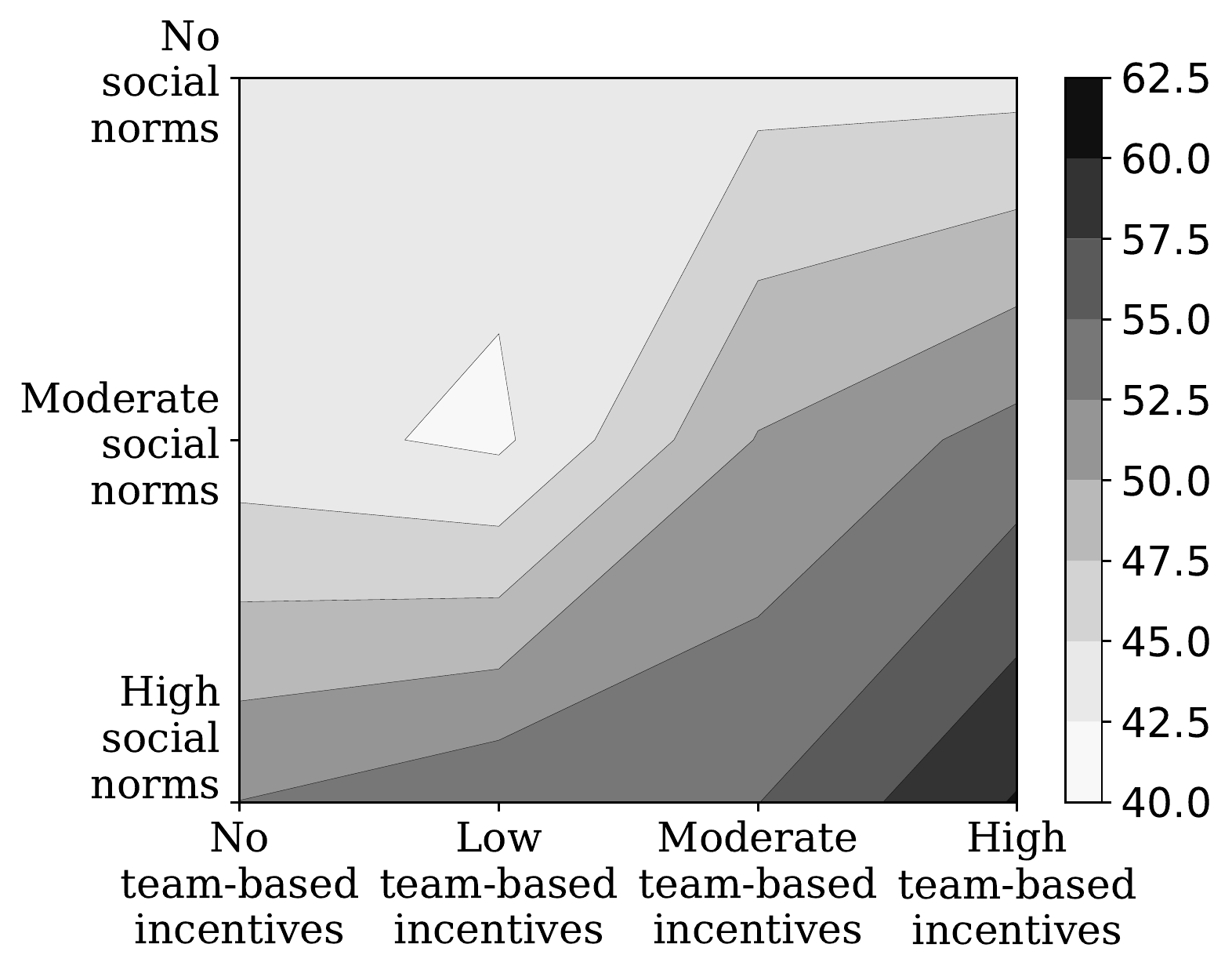}
        \captionsetup{justification=centering}
		\caption*{(a) Internal complexity\\{\small ($K=3,C=S=0$)}}
	\end{minipage}
	\begin{minipage}{0.49\linewidth}
		\includegraphics[width=\linewidth]{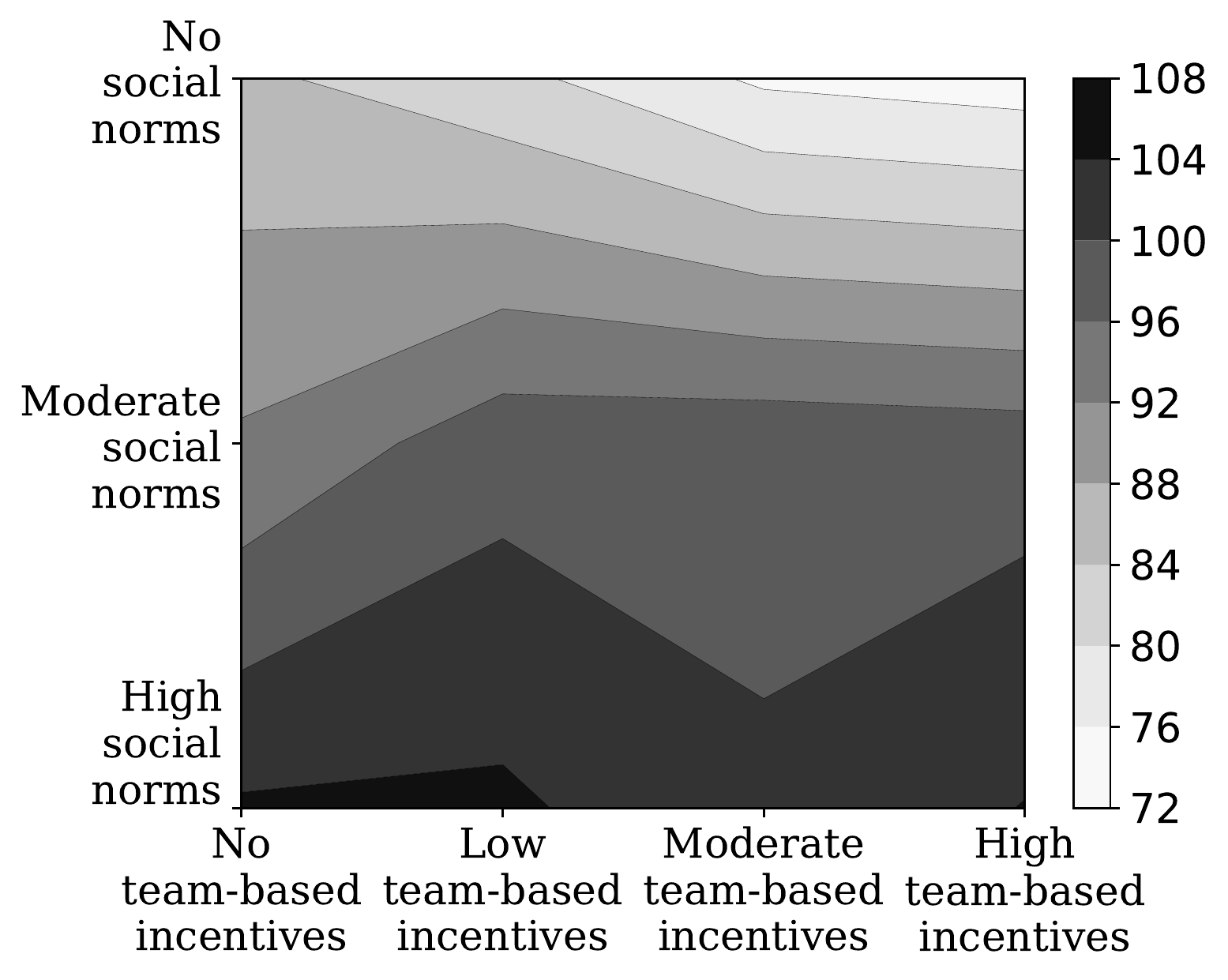}
        \captionsetup{justification=centering}
		\caption*{(b) Low complexity\\{\small ($K=C=S=1$)}}
	\end{minipage}
	\\
	\begin{minipage}{0.49\linewidth}
		\includegraphics[width=\linewidth]{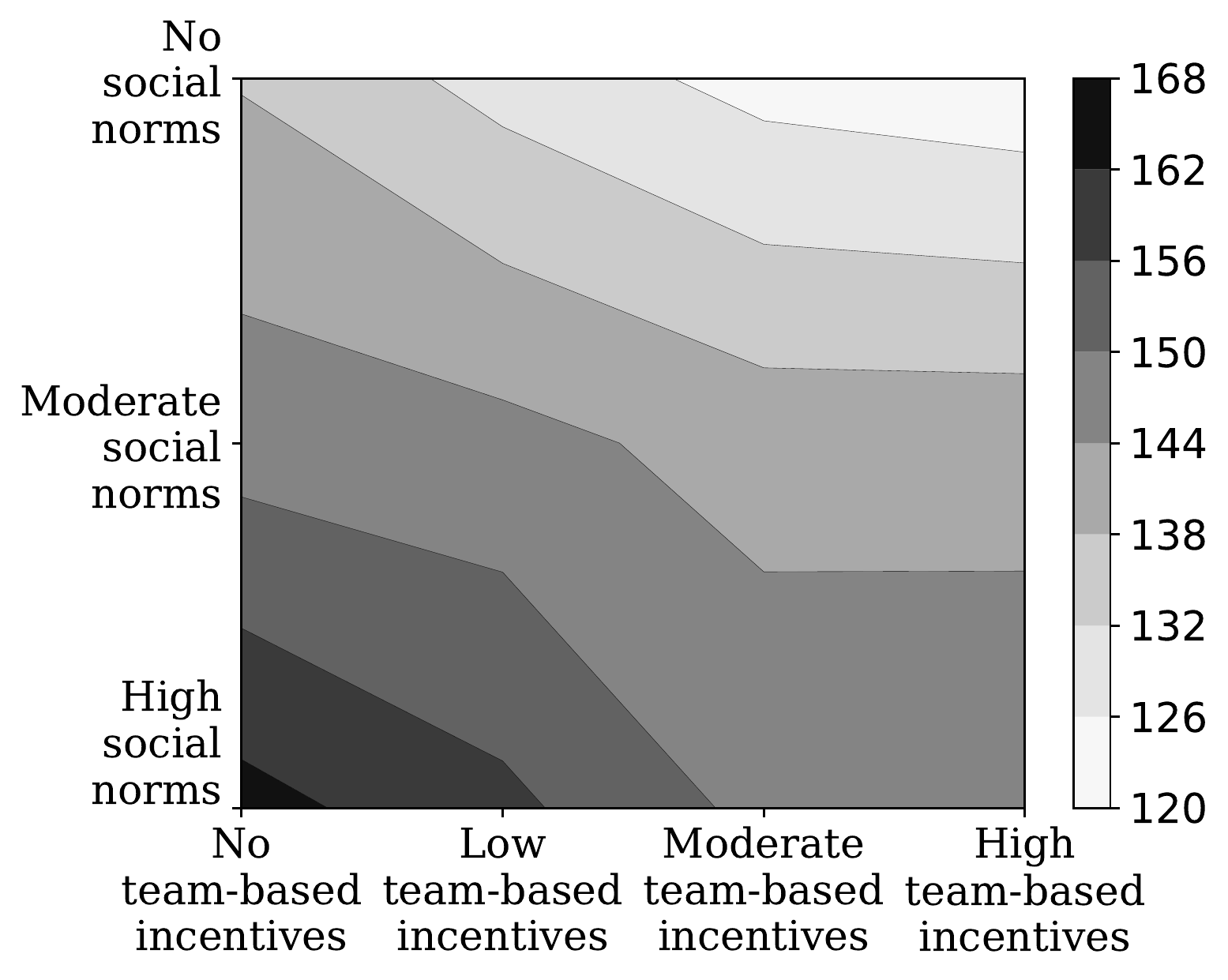}
        \captionsetup{justification=centering}
		\caption*{(c) Moderate complexity\\{\small ($K=C=S=2$)}}
	\end{minipage}
	\begin{minipage}{0.49\linewidth}
		\includegraphics[width=\linewidth]{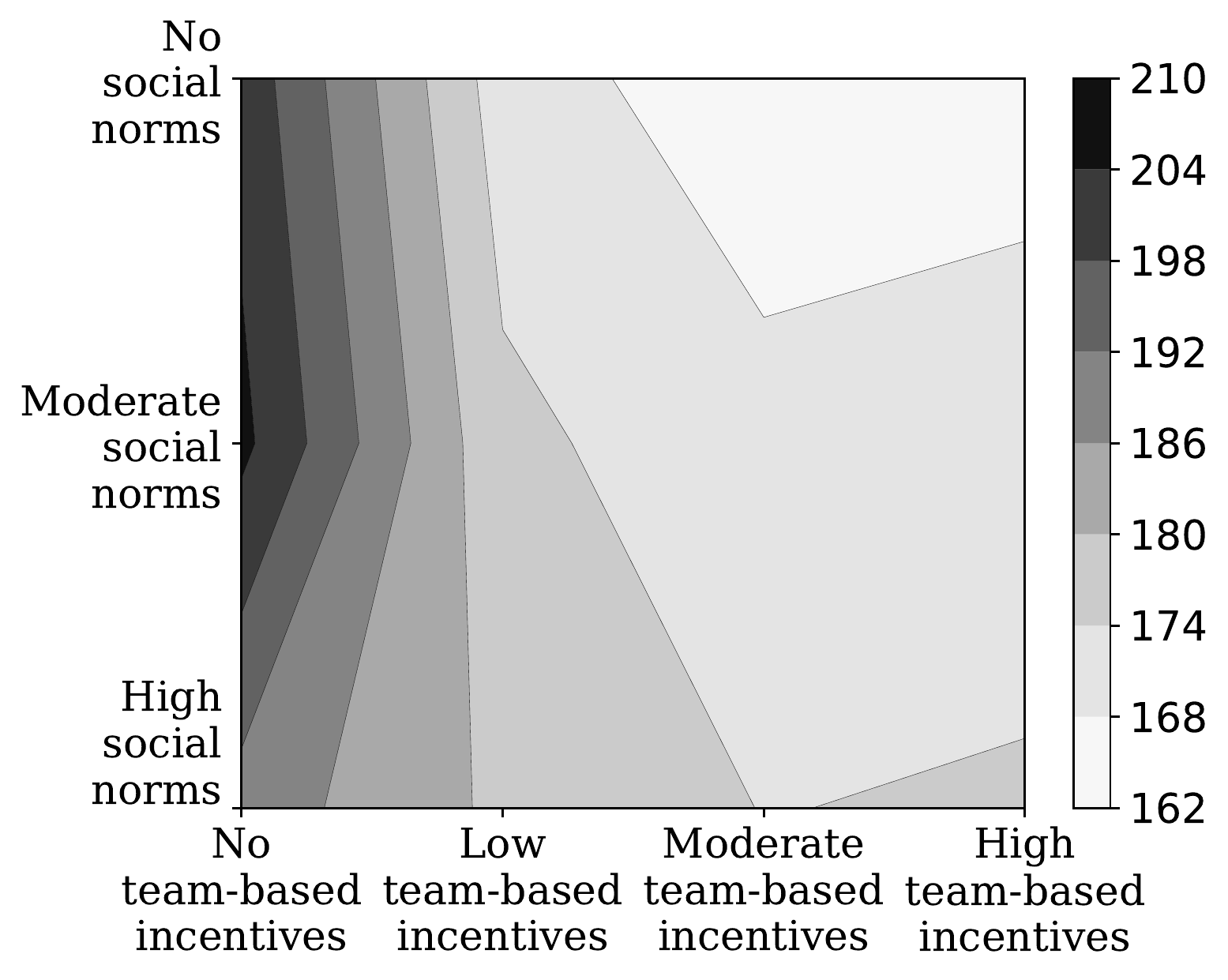}
        \captionsetup{justification=centering}
		\caption*{(d) High complexity\\{\small ($K=S=3,C=4$)}}
	\end{minipage}
	\caption{Contour plots for cumulative distances $d(\overline{\Phi})$ to the maximum attainable performance for different scenarios. The lower (higher) values mean better (worse) performance for organization and are indicated by lighter (darker) tones}
	\label{fig:results}
\end{figure}

The parameters summarized in Tab. \ref{tab:params} result in $4 \cdot 3 \cdot 4 = 48$ different scenarios for 4 levels of complexity (internal, low, moderate, and high), 3 pairs of weights for performance-based incentive and compliance with social norms (high, moderate, and zero weight on social norms), and 4 different settings for the performance-based incentive schemes (zero, low, moderate, and high team-based incentives). 

The results are presented in Fig. \ref{fig:results}. The contours indicate ranges of similar distance values, where darker (lighter) colors indicate larger (smaller) values for the distance to maximum. In other words, the lighter contours represent higher organizational performance and are more desirable, while the darker contours represent lower organizational performance and are less desirable. In each plot, the performance-based incentive scheme ($\alpha$ and $\beta$) and the pairs of weights for incentives and social norms ($w_{inc}$ and $w_{soc}$) are presented as the horizontal and the vertical axes, respectively. Please note that performance-based incentive schemes that put full weight on the performance achieved by the agents individually are included on the left hand side on horizontal axes (i.e., $\alpha=1$ and $\beta=0$), while moving to the right decreases the weight of individual performance and increases the weight of residual performance (until $\alpha=0.25$ and $\beta=0.75$). On the vertical axes, scenarios in which agents put strong emphasis on complying with social norms, are included at the bottom (i.e., $w_{inc}=0.5$ and $w_{soc}=0.5$), while moving upwards  decreases the extent to which agents care about social norms (until $w_{inc}$ = 1 and $w_{soc}=0$). The 4 contour plots correspond to 4 different levels  of complexity presented in Fig. \ref{fig:interactions}.

Looking at the ranges of the plots (the minimal and maximal values), we observe that as the complexity increases, the average performance drops for all social norm weights and incentive schemes. This finding is in line with previous research \cite{kauffman91,levinthal97}.

For scenarios in which agents put full emphasis on performance-based incentives and do not care about complying with social norms (i.e., upper parts of subplots, where $w_{inc}=1$ and $w_{soc}=0$), we can observe that the choice of the incentive scheme does not have an effect on performance in the absence of external interdependencies (see Fig. \ref{fig:results} (a)). However as soon as there are external interdependencies, even if the task's complexity is relatively low (see Fig. \ref{fig:results} (b)), the team-based incentive schemes result in a better performance.\footnote{Please note that task complexity in Fig. \ref{fig:results} (b) is relatively low, since  every task is coupled with $K+C \cdot S = 2$ other tasks. In Fig.  \ref{fig:results} (a), on the contrary, each task is coupled with $K + C \cdot S = 3$ other tasks.} 
This positive effect of team-based incentive mechanisms increases with the external complexity of the task environment (see Fig. \ref{fig:results} (b,c,d)). This finding emphasizes the importance of differentiating between internal and external interdependencies among tasks when designing incentive mechanisms. A stronger focus of incentives on the residual performance (higher values of $\beta$) appears to offset some of the negative effects associated with task complexity only in cases in which the complexity is not internal (i.e., when $C,S>0$). This finding, therefore, extends the literature that states that residual performance is suitable in presence of externalities \cite{fischer08}, by specifying the nature of these externalities. We also find that the pattern described above is robust in presence of social norms (i.e., on lower parts of subplots, the similar effect is observed).

We also observe that as the agents start putting higher weights on the social norms (i.e. moving down the vertical axis), the contours get darker, meaning that the performance drops. This represents that complying to social norms can come at a cost for performance, as agents have to consider multiple objectives. However, as the (external) complexity increases (Fig. \ref{fig:results} (b,c,d)), the extent of this effect declines, and in situations with high (external) complexity (see Fig. \ref{fig:results} (d)) we observe that the contours are almost vertical, meaning that social norms do not cause a significant decline in the performance. This can be explained by the coordinating function of social norms, which can be observed when the task environment is too complex to solve individually without any coordination. Apart from that, our sensitivity analyses show that the decline in performance related to social norms, disappears for cases with higher correlation among agents' performance landscapes even for environments with lower complexity. 
\section{CONCLUSION}
\label{sec:conclusion}

In this paper, we proposed a model of an organization which is composed of autonomous and collaborative decision making agents facing a complex task. Agents pursue two objectives simultaneously, i.e., they aim at maximizing their performance-based incentives and, at the same time, want to comply to the social norms emerging in their social networks. In our analysis, we focus on the interplay between performance-based incentives and social norms. Our main results are the following:
First, if agents focus on performance-based incentives only, the choice of the type of incentive scheme has marginal effects in task environments with low level of complexity. As complexity increases, team-based incentives become more beneficial. However, in environments where inter-dependencies (no matter how high) exist only within tasks allocated to the same agent, the incentive schemes have zero effect on the performance.
Second, if agents focus on complying to social norms, this comes at the cost of performance at the level of the system (except for scenarios when agents' task environments are highly correlated).
Third, whether team-based performance can offset negative effects on performance, caused by agents that aim at complying to social norms, is substantially affected by the level of task complexity. For highly complex tasks, team-based incentives appear to be beneficial, while the opposite is true for task environments with a low level of complexity. 

Our work is, of course, not without its limitations. First, we treat compliance and non-compliance to social norms equally. In reality, however, non-compliance to social norms might lead to more fatal consequences than ``over-compliance'' \cite{festre10}. Future work might want to investigate this issue. Second, we limit the number of agents to $4$ and consider ring networks only. It might be a promising avenue for future research to increase the number of agents and test the effect of other network topologies on the dynamics emerging from social norms. Finally, it might be an interesting extension to model the transformation of social norms into values by adjusting the task environment dynamically.

\bibliographystyle{splncs04}
\bibliography{refs.bib}
\end{document}